\documentclass[12pt]{article}
\usepackage[lmargin=.75in,rmargin=.75in,tmargin=1.in,bmargin=1in]{geometry}
\usepackage{psfrag}
\usepackage{graphicx}
\usepackage{graphics}

\newcommand{\as}{a\!\!\!/}
\newcommand{\As}{A\!\!\!/}
\newcommand{\ks}{k\!\!\!/}
\newcommand{\ps}{p\!\!\!/}
\newcommand{\ets}{\eta\!\!\!/}

\newcommand{\oll}{\overline}

\date{}
\begin{document}
\title{Mott scattering in strong laser fields revisited.}
\author{B. Manaut$^{\ddag\dag}$\thanks{manaut\_bouzid@yahoo.fr}, Y.  Attaourti$^{\dag}$ , S. Taj$^\dag$,  and S. Elhandi  $^\dag$\\
 {\it {\small $^\dag$ Laboratoire de Physique des Hautes
Energies et d'Astrophysique, Facult\'e des Sciences}}\\
{\it {\small  Semlalia, Universit\'e Cadi Ayyad Marrakech, BP :
2390, Maroc.}}
\\
{\it {\small $^\ddag$ L.I.R.S.T, Facult\'e polydisciplinaire, Universit\'e Sultan Moulay Slimane B\'eni Mellal, BP : 523, Maroc.}}\\
 \\
  }
\maketitle
\begin{abstract}
In this work, we review and correct the first Born differential
cross section for the process of Mott scattering of a Dirac-Volkov
electron, namely, the expression (26) derived by Szymanowski et
al [Physical Review A {\bf 56}, 3846 (1997)]. In particular, we
disagree with the expression of
$\left(\frac{d\sigma}{d\Omega}\right)$ they obtained and we give
the exact coefficients multiplying the various Bessel functions
appearing in the scattering differential cross section.
Comparison of our numerical calculations with those of
Szymanowski et al. shows qualitative and quantitative differences
when the incoming total electron energy and the electric field
strength are increased particularly in the direction of the laser
propagation. Such corrections are very important since the
relativistic electronic dressing of any Dirac-Volkov charged
particle gives rise to these coefficients that multiply the
various Bessel functions and the relativistic study of other
processes (such as excitation, ionisation, etc....) depends
strongly of the correctness and reliability of the calculations
for this process of Mott Scattering in presence of a laser field.
Our work has been accepted [Y. Attaourti, B. Manaut, Physical
Review A {\bf 68}, 067401 (2003)] but only as a comment. In this
paper, we give the full details of the calculations as well as
the clear explanation of the large discrepancies that their
results could cause when working in the ultra relativistic regime
and using a very strong laser field corresponding to an electric
field $\varepsilon=5.89$ in atomic units.
  \\
\vspace{.04cm}\\
 PACS number(s): 34.80.Qb, 12.20.Ds
\end{abstract}
\section{Introduction}
In a pioneering and very often cited paper , Szymanowski et al.
\cite{1} have studied the Mott scattering process in a strong
laser field. The main purpose of their work was to show that the
modifications of the Mott scattering differential cross section
for the scattering of an electron by the Coulomb potential of a
nucleus in the presence of a strong laser field, can yield
interesting physical insights concerning the importance and the
signatures of the relativistic effects. Their spin dependent
relativistic description of Mott scattering permits to
distinguish between kinematics and spin-orbit coupling effects.
They have compared the results of a calculation of the first Born
differential cross section for the Coulomb scattering of the
Dirac-Volkov electrons dressed by a circularly polarized laser
field to the first Born cross section for the Coulomb scattering
of spinless Klein-Gordon particles and also to the non
relativistic Schr\"odinger-Volkov treatment. The aim of our work
is to provide the correct expression for the first-Born
differential cross sections corresponding to the Coulomb
scattering of the Dirac-Volkov electrons. On the one hand, we
show that the terms proportional to $ sin(2\phi_0)$ are missing
in \cite{1}, where $\phi_0$ is the phase stemming from the
expression of the circularly polarized electromagnetic field. The
claim of \cite{1} that they vanish is not true. These terms do
not depend on the chosen description of the circular polarization
in cartesian components. On the other hand, we perform the
calculations with some details and throughout this work, we use
atomic units $(\hbar=e=m=1)$ where $m$ denotes the electron mass.
The abbreviation DCS stands for the differential cross
section.\\
\indent The organization of this paper is as follows : in Section
2, we establish the expression of the $S$-matrix transition
amplitude as well as the formal expression of scattering DCS. In
Section 3, we give a detailed account on the various trace
calculations and show that indeed there is a missing term
proportional to $\sin(2\phi_0)$ that is not equal to zero. This
term as well as a term proportional to $\cos(2\phi_0)$ contribute
to $\left(\frac{d\sigma}{d\Omega}\right)$ and multiply the product
$J_{s+1}(z)J_{s-1}(z)$, where $J_{s}(z)$ is an ordinary Bessel
function of argument $z$ and index $s$. The argument $z$
appearing in the above mentioned product will be defined later.
Then, we carry out the derivation of the correct expression of
the scattering DCS associated to the exchange of a given number
of laser photons. In section 4, we give some estimates of the
numerical significance of our corrections. In particular, we
compare numerically the Dirac-Volkov DCS we have obtained with
the corresponding DCS of \cite{1}. We end by a brief conclusion
in Section 5.
\section{The $S$-matrix element and the scattering differential
cross section}

\noindent Exact solutions of relativistic wave equations \cite{2}
are very difficult to obtain. However, in a seminal paper, Volkov
\cite{3} obtained the formal solution of the Dirac equation for
the relativistic electron with 4-momentum $p^{\mu}$ inside a
classical monochromatic electromagnetic field $A^{\mu}$. These
solutions are called the relativistic Volkov states. The plane
wave electromagnetic field $A^{\mu}$ of 4-momentum $k^{\mu}$
 $(k_{\mu}k^{\mu}=k^2=0)$ depends only on the argument
 $\phi=k.x=k_{\mu}x^{\mu}$ and therefore $A^{\mu}$ is such that
 \begin{equation}
A^{\mu}=A^{\mu}(k.x)=A^{\mu}(\phi). \label{1}
 \end{equation}
 The 4-vector $A^{\mu}$ satisfies the Lorentz gauge condition
 $\partial_{\mu}A^{\mu}=0$ or equivalently $k_{\mu}A^{\mu}=0$. The
 Dirac-Volkov equation for an electron in an external field $A_{\mu}$ is
\begin{equation}
\left\{(\hat{p}-\frac{1}{c}A)^2-c^2-\frac{i}{2c}F_{\mu\nu}\sigma^{\mu\nu}\right\}\psi(x)=0,
\label{2}
 \end{equation}
 where $F_{\mu\nu}$ is the electromagnetic field tensor
 $F_{\mu\nu}=\partial_{\mu}A_{\nu}-\partial_{\nu}A_{\mu}$ and
 $\sigma^{\mu\nu}=\frac{1}{2}[\gamma^{\mu},\gamma^{\nu}]$. The
 matrices $\gamma^{\mu}$ are the anticommuting Dirac matrices such
 that
 $\gamma^{\mu}\gamma^{\nu}+\gamma^{\nu}\gamma^{\mu}=2g^{\mu\nu}1_4$,
 where $g^{\mu\nu}$ is the metric tensor
 $g^{\mu\nu}=diag(1,-1,-1,-1)$ and $1_4$ is the identity matrix in
 four dimensions. The solutions of Eq.(\ref{2}) are the
 relativistic Dirac-Volkov wave functions
 \begin{equation}
\psi_p(x)=R(p)\frac{u(p,s)}{\sqrt{2p_0V}}e^{iS(x)}, \label{3}
 \end{equation}
 whith
 \begin{equation}
R(p)=\exp{(\frac{\ks\As}{2c(k.p)})}=1+\frac{\ks\As}{2c(k.p)},
\label{4}
 \end{equation}
 and the function $S(x)$ is given by
\begin{equation}
S(x)=-p.x-\int_{0}^{k.x}\frac{1}{c(k.p)}\left[p.A(\xi)-\frac{1}{2c}A^2(\xi)\right]d\xi.
\label{5}
 \end{equation}
 In Eq.(\ref{3}), $u(p,s)$ represents a Dirac bispinor which
 satisfies the free Dirac equation and is normalized according to
 $\oll{u}(p,s)u(p,s)=u^*(p,s)\gamma^0u(p,s)=2c^2$. We consider a
 circularly polarized field
 \begin{equation}
A=a_1\cos(\phi)+a_2\sin(\phi), \label{6}
 \end{equation}
 where $\phi=k.x$. We choose $a_1^2=a_2^2=a^2=A^2$ and
 $a_1.a_2=a_2.a_1=0$. The Lorentz condition $k.A=0$ implies
 $a_1.k=a_2.k=0$. If one assumes that $A^{\mu}$ is
 quasi-periodic so that its time average is zero $\oll{A^{\mu}}=0$,
 then using the Gordon identity, the averaged 4-current is easily
 obtained :
\begin{equation}
\oll{j^{\mu}}=\frac{1}{p_0}\left\{p^{\mu}-\frac{1}{2c^2(k.p)}\oll{A^2}k^{\mu}\right\}.
\label{7}
 \end{equation}
 If one sets
\begin{equation}
q^{\mu}=p^{\mu}-\frac{1}{2c^2(k.p)}\oll{A^2}k^{\mu}, \label{8}
 \end{equation}
 this yields
\begin{equation}
q.q=q^{\mu}q_{\mu}=m_*^2c^2, \label{9}
 \end{equation}
 with
\begin{equation}
m_*^2=1-\frac{\oll{A^2}}{c^4}. \label{10}
 \end{equation}
 One often calls the averaged 4-momentum $q^{\mu}$ a
 quasi-impulsion. Note that $q^{\mu}=(Q/c,\mathbf{q})$. The quantity
 $m_*$ plays the role of an effective mass of the electron inside
 the electromagnetic field. For the study of the process of Mott
 scattering in presence of a laser field, we use the Dirac-Volkov
 wave functions \cite{3} normalized in the volume $V$:
 \begin{equation}
\psi_q(x)=R(p)\frac{u(p,s)}{\sqrt{2QV}}e^{iS(q,x)}, \label{11}
 \end{equation}
 whith
\begin{eqnarray}
R(p)&=&R(q)=1+\frac{1}{2c(k.p)}\ks\As\nonumber\\
&=&1+\frac{1}{2c(k.p)}(\ks.\as_1\cos(\phi)+\ks.\as_2\sin(\phi))\label{12}
 \end{eqnarray}
 and
 \begin{eqnarray}
S(q,x)&=&-q.x-\frac{(a_1.p)}{c(k.p)}\sin(\phi)+\frac{(a_2.p)}{c(k.p)}\cos(\phi)\nonumber\\
&=&-q.x-\frac{(a_1.q)}{c(k.q)}\sin(\phi)+\frac{(a_2.q)}{c(k.q)}\cos(\phi).
\label{13}
 \end{eqnarray}
 We turn now to the calculation of the transition amplitude. The
 interaction of the dressed electrons with the central Coulomb
 field
\begin{equation}
A^{\mu}=(-\frac{Z}{|\mathbf{x}|},0,0,0)\label{14}
 \end{equation}
 is considered as a first-order perturbation. This is well
 justified if $Z\alpha \ll 1$, where $Z$ is the nuclear charge of
 the nucleus considered and $\alpha$ is the fine-structure
 constant. We evaluate the transition matrix element for the
 transition ($i\rightarrow f$)
\begin{equation}
S_{fi}=\frac{iZ}{c}\int d^4x
\oll{\psi}_{qf}(x)\frac{\gamma^0}{|\mathbf{x}|}\psi_{qi}(x).
\label{15}
 \end{equation}
 We first consider the quantity
\begin{equation}
\oll{\psi}_{qf}(x)\frac{\gamma^0}{|\mathbf{x}|}\psi_{qi}(x)=\frac{1}{\sqrt{2Q_i
V }}\frac{1}{\sqrt{2Q_f V
}}\oll{u}(p_f,s_f)\oll{R}(p_f)\frac{\gamma^0}{|\mathbf{x}|}R(p_i)u(p_i,s_i)e^{-i(S(q_f,x)-S(q_i,x))}.
\label{16}
 \end{equation}
 We have
\begin{equation}
e^{-i(S(q_f,x)-S(q_i,x))}=\exp[i(q_f-q_i).x-iz\sin(\phi-\phi_0)],
\label{17}
 \end{equation}
 where $z$ is such that
\begin{equation}
z=\sqrt{\alpha_1^2+\alpha_2^2}, \label{18}
 \end{equation}
 whereas the quantities $\alpha_1$ and $\alpha_2$ are given by
\begin{equation}
\alpha_1=\frac{(a_1.p_i)}{c(k.p_i)}-\frac{(a_1.p_f)}{c(k.p_f)},\,
\alpha_2=\frac{(a_2.p_i)}{c(k.p_i)}-\frac{(a_2.p_f)}{c(k.p_f)}
\label{19}
 \end{equation}
 and the phase $\phi_0$ is such that
 $\phi_0=\arccos(\alpha_1/z)=\arcsin(\alpha_2/z)=\arctan(\alpha_2/\alpha_1)$.
 It is important at this stage to perform intermediate
 calculations in order to reduce the numbers of $\gamma$ matrices
 that will appear when one calculates the scattering DCS. After
 some algebraic manipulations, one gets
\begin{eqnarray}
&&\oll{u}(p_f,s_f)\oll{R}(p_f)\gamma^0R(p_i)u(p_i,s_i)\nonumber\\
&=&\oll{u}(p_f,s_f)[C_0+C_1\cos(\phi)+C_2\sin(\phi)]u(p_i,s_i),
\label{20}
 \end{eqnarray}
 where the three coefficients $C_0$, $C_1$ and $C_2$ are
 respectively given by
\begin{eqnarray}
C_0&=&\gamma^0-2k_0a^2\ks c(p_i)c(p_f)\nonumber\\
C_1&=&c(p_i)\gamma^0\ks\as_1+c(p_f)\as_1\ks\gamma^0\nonumber\\
C_2&=&c(p_i)\gamma^0\ks\as_2+c(p_f)\as_2\ks\gamma^0\label{21}
 \end{eqnarray}
 with $c(p)=\frac{1}{2c(k.p)}$ and $k_0=k^0=\omega/c$. Therefore,
 the transition matrix element becomes
\begin{eqnarray}
S_{fi}&=&\frac{iZ}{c}\int
d^4x\frac{1}{\sqrt{2Q_iV}}\frac{1}{\sqrt{2Q_fV}}
\oll{u}(p_f,s_f)[C_0+C_1\cos(\phi)+C_2\sin(\phi)]u(p_i,s_i)\nonumber\\
&\times&\exp[i(q_f-q_i).x-izsin(\phi-\phi_0)]. \label{22}
 \end{eqnarray}
 We now invoke the well-known identities involving ordinary Bessel
 functions $J_s(z)$
\begin{eqnarray}
\left\{\begin{array}{c}
1\\
\cos(\phi)\\
\sin(\phi)\end{array}\right\}e^{-iz\sin(\phi-\phi_0)}=\sum_{s=-\infty}^{\infty}\left\{\begin{array}{c}
B_s\\
B_{1s}\\
B_{2s}\end{array}\right\}e^{-is\phi},\label{23}
\end{eqnarray}
with
\begin{eqnarray}
\left\{\begin{array}{c}
B_s\\
B_{1s}\\
B_{2s}\end{array}\right\}=\left\{\begin{array}{c}
J_s(z)e^{is\phi_0}\\
(J_{s+1}(z)e^{i(s+1)\phi_0}+J_{s-1}(z)e^{i(s-1)\phi_0})/2\\
(J_{s+1}(z)e^{i(s+1)\phi_0}-J_{s-1}(z)e^{i(s-1)\phi_0})/2i\end{array}\right\}.
\label{24}
\end{eqnarray}
Evaluating the integrals over $x_0$ and $\mathbf{x}$ yields for
$S_{fi}$ :
\begin{equation}
S_{fi}=\frac{i4\pi
Z}{\sqrt{2Q_iV}\sqrt{2Q_fV}}\sum_{s=-\infty}^{\infty}\frac{2\pi\delta(Q_f-Q_i-s\omega)
}{|\mathbf{q}_f-\mathbf{q}_i-s\mathbf{k}|^2}M_{fi}^{(s)},\label{25}
\end{equation}
where the quantity $M_{fi}^{(s)}$ is defined by
\begin{equation}
M_{fi}^{(s)}=\oll{u}(p_f,s_f)[C_0B_s+C_1B_{1s}+C_2B_{2s}]u(p_i,s_i).
\label{26}
\end{equation}
To evaluate the DCS, we first evaluate the transition probability
per particle into final states within the range of momentum
$d\mathbf{q}_f$
\begin{eqnarray}
dW_{fi}&=&|S_{fi}|^2\frac{Vd\mathbf{q}_f}{(2\pi)^3}\nonumber\\
&=&\frac{(4\pi)^2Z^2}{2Q_iV.2Q_fV}\sum_{s=-\infty}^{\infty}\frac{T2\pi
\delta(Q_f-Q_i-sw)}{|\mathbf{q}_f-\mathbf{q}_i-s\mathbf{k}|^4}|M_{fi}^{(s)}|^2\frac{Vd\mathbf{q}_f}{(2\pi)^3},
\label{27}
\end{eqnarray}
where we have used the rule of replacement
\begin{eqnarray}
[2\pi\delta(Q_f-Q_i-sw)]^2& \rightarrow
&2\pi\delta(0)2\pi\delta(Q_f-Q_i-sw)\nonumber\\
&=&T2\pi\delta(Q_f-Q_i-sw). \label{28}
\end{eqnarray}
Next, we have for the transition probability per unit time
\begin{eqnarray}
dR_{fi}&=&\frac{dW_{fi}}{T}\nonumber\\
&=&\frac{(4\pi)^2Z^2}{2Q_iV.2Q_fV}\sum_{s=-\infty}^{\infty}\frac{2\pi
\delta(Q_f-Q_i-sw)}{|\mathbf{q}_f-\mathbf{q}_i-s\mathbf{k}|^4}|M_{fi}^{(s)}|^2\frac{Vd\mathbf{q}_f}{(2\pi)^3}.
\label{29}
\end{eqnarray}
Dividing $dR_{fi}$ by the flux of incoming particles
\begin{equation}
|\mathbf{J}^{inc}|=\frac{|\mathbf{q}_i|c^2}{Q_iV}, \label{30}
\end{equation}
then using the relation
$|\mathbf{q}_f|d|\mathbf{q}_f|=\frac{1}{c^2}Q_fdQ_f$ and
integrating over the final energy, we get for the scattering DCS
\begin{eqnarray}
\frac{d\sigma}{d\Omega_f}&=&\left.\frac{Z^2}{c^4}\frac{|\mathbf{q}_f|}{|\mathbf{q}_i|}\sum_{s=\-\infty}^{\infty}\frac{|M_{fi}^{(s)}|^2}{|\mathbf{q}_f-\mathbf{q}_i-s\mathbf{k}|^4}\right|_{Q_f=Q_i+sw}\nonumber\\
&=&\left.\sum_{s=-\infty}^{\infty}\frac{d\sigma^{(s)}}{d\Omega_f}\right|_{Q_f=Q_i+sw},
\label{31}
\end{eqnarray}

where
\begin{equation}
 \left.\frac{d\sigma^{(s)}}{d\Omega_f}\right|_{Q_f=Q_i+sw}=\left.\frac{Z^2}{c^4}\frac{|\mathbf{q}_f|}{|\mathbf{q}_i|}\frac{|M_{fi}^{(s)}|^2}{|\mathbf{q}_f-\mathbf{q}_i-s\mathbf{k}|^4}\right|_{Q_f=Q_i+sw}. \label{32}
\end{equation}
The calculation is now reduced to the computation of traces of
$\gamma$ matrices. This is routinely done using Reduce \cite{4}.
We consider the unpolarized DCS. Therefore, the various
polarization states have the same probability and the actually
measured DCS is given by summing over the final polarization $s_f$
and averaging over the initial polarization $s_i$. Therefore, the
unpolarized DCS is formally given by
\begin{equation}
 \frac{d\oll{\sigma}}{d\Omega_f}=\left.\sum_{s=-\infty}^{\infty}\frac{d\oll{\sigma}^{(s)}}{d\Omega_f}\right|_{Q_f=Q_i+sw}, \label{33}
\end{equation}
where
\begin{eqnarray}
\left.\frac{d\oll{\sigma}^{(s)}}{d\Omega_f}\right|_{Q_f=Q_i+sw}&=&\frac{Z^2}{c^4}\frac{|\mathbf{q}_f|}{|\mathbf{q}_i|}\frac{1}{|\mathbf{q}_f-\mathbf{q}_i-s\mathbf{k}|^4}\left.\frac{1}{2}\sum_{s_i}\sum_{s_f}|M_{fi}^{(s)}|^2\right|_{Q_f=Q_i+sw}.
\label{34}
\end{eqnarray}
\section{Trace calculations.}
Since the controversy is very acute and precise about the results
of the sum over the polarization
$\frac{1}{2}\sum_{s_i}\sum_{s_f}|M_{fi}^{(s)}|^2$, we devote a
whole section to the calculations of the various traces that
intervene in the formal expression of the unpolarized DCS given
by Eq.(\ref{34}). We have to calculate
\begin{eqnarray}
\frac{1}{2}\sum_{s_i}\sum_{s_f}|M_{fi}^{(s)}|^2&=&\frac{1}{2}\sum_{s_i}\sum_{s_f}|\oll{u}(p_f,s_f)[C_0B_s+C_1B_{1s}+C_2B_{2s}]u(p_i,s_i)|^2\nonumber\\
&=&\frac{1}{2}\sum_{s_i}\sum_{s_f}|\oll{u}(p_f,s_f)\Lambda^{(s)}u(p_i,s_i)|^2,
\label{35}
\end{eqnarray}
with
\begin{eqnarray}
\Lambda^{(s)}&=&[\gamma^0-2k_0a^2\ks c(p_i)c(p_f)]B_s\nonumber\\
&+&[c(p_i)\gamma^0\ks\as_1+c(p_f)\as_1\ks\gamma^0]B_{1s}\nonumber\\
&+&[c(p_i)\gamma^0\ks\as_2+c(p_f)\as_2\ks\gamma^0]B_{2s}.
\label{36}
\end{eqnarray}
Using standard techniques of the $\gamma$ matrix algebra, one has
 \begin{equation}
\frac{1}{2}\sum_{s_i}\sum_{s_f}|M_{fi}^{(s)}|^2=\frac{1}{2}Tr\{(\ps_fc+c^2)\Lambda^{(s)}(\ps_ic+c^2)\oll{\Lambda}^{(s)}\},
\label{37}
\end{equation}
with
\begin{eqnarray}
\oll{\Lambda}^{(s)}&=&\gamma^0\Lambda^{(s)\dag}\gamma^0\nonumber\\
&=&[\gamma^0-2k_0a^2\ks c(p_i)c(p_f)]B^*_s\nonumber\\
&+&[c(p_i)\as_1\ks\gamma^0+c(p_f)\gamma^0\ks\as_1]B^*_{1s}\nonumber\\
&+&[c(p_i)\as_2\ks\gamma^0+c(p_f)\gamma^0\ks\as_2]B^*_{2s}.
\label{38}
\end{eqnarray}
There are nine main traces to be calculated. We write them
explicitly
\begin{eqnarray}
&&\mathcal{M}_1=Tr\{(\ps_fc+c^2)C_0(\ps_ic+c^2)\oll{C}_0\}|B_s|^2,\nonumber\\
&&\mathcal{M}_2=Tr\{(\ps_fc+c^2)C_0(\ps_ic+c^2)\oll{C}_1\}B_sB^*_{1s},\nonumber\\
&&\mathcal{M}_3=Tr\{(\ps_fc+c^2)C_0(\ps_ic+c^2)\oll{C}_2\}B_sB^*_{2s},\nonumber\\
&&\mathcal{M}_4=Tr\{(\ps_fc+c^2)C_1(\ps_ic+c^2)\oll{C}_0\}B^*_sB_{1s},\nonumber\\
&&\mathcal{M}_5=Tr\{(\ps_fc+c^2)C_1(\ps_ic+c^2)\oll{C}_1\}|B_{1s}|^2, \label{39}\\
&&\mathcal{M}_6=Tr\{(\ps_fc+c^2)C_1(\ps_ic+c^2)\oll{C}_2\}B_{1s}B^*_{2s},\nonumber\\
&&\mathcal{M}_7=Tr\{(\ps_fc+c^2)C_2(\ps_ic+c^2)\oll{C}_0\}B_{2s}B^*_s,\nonumber\\
&&\mathcal{M}_8=Tr\{(\ps_fc+c^2)C_2(\ps_ic+c^2)\oll{C}_1\}B^*_{1s}B_{2s},\nonumber\\
&&\mathcal{M}_9=Tr\{(\ps_fc+c^2)C_2(\ps_ic+c^2)\oll{C}_2\}|B_{2s}|^2.\nonumber
\end{eqnarray}
To simplify the notations, we will drop the argument of the
various ordinary Bessel functions that appear. The diagonal terms
give rise to
\begin{eqnarray}\left.\begin{array}{c}
\mathcal{M}_1\propto|B_s|^2=J_s^2,\\
\mathcal{M}_5\propto|B_{1s}|^2=\frac{1}{4}(J^2_{s+1}+2J_{s+1}J_{s-1}\cos(2\phi_0)+J^2_{s-1}),\\
\mathcal{M}_9\propto|B_{2s}|^2=\frac{1}{4}(J^2_{s+1}-2J_{s+1}J_{s-1}\cos(2\phi_0)+J^2_{s-1}).
\label{40}
\end{array}\right.
\end{eqnarray}
So, taking into account the fact that the traces multiplying
$|B_s|^2$, $|B_{1s}|^2$ and $|B_{2s}|^2$ are not zero, one
expects that terms proportional to $J_{s+1}J_{s-1}\cos(2\phi_0)$
will be present in the expression of the scattering DCS. The
first controversy between our work and the result of Szymanowski
et al \cite{1} concerns the traces $\mathcal{M}_6$
and$\mathcal{M}_8$. Since
\begin{eqnarray}\left.\begin{array}{c}
\mathcal{M}_6\propto B_{1s}B^*_{2s}=\frac{i}{4}(J^2_{s+1}-2iJ_{s+1}J_{s-1}\sin(2\phi_0)-J^2_{s-1})\\
\mathcal{M}_8\propto B_{1s}^*
B_{2s}=\frac{-i}{4}(J^2_{s+1}+2iJ_{s+1}J_{s-1}\sin(2\phi_0)-J^2_{s-1})
\end{array}\right. \label{41}
\end{eqnarray}
and with little familiarity with the $\gamma$ matrix algebra, one
can see at once that if the corresponding traces are not zero
then the net contribution of $\mathcal{M}_6+\mathcal{M}_8$ will
contain a term proportional to $J_{s+1}J_{s-1}\sin(2\phi_0)$. We
shall demonstrate that in what follows. We have
\begin{eqnarray}
\mathcal{M}_6&=&Tr\{(\ps_f c+c^2)C_1(\ps_i
c+c^2)\oll{C}_2\}B_{1s}B_{2s}^*\nonumber\\
&=&Tr\{(\ps_f
c+c^2)[c(p_i)\gamma^0\ks\as_1+c(p_f)\as_1\ks\gamma^0](\ps_i
c+c^2)\nonumber\\
&&[c(p_i)\as_2\ks\gamma^0+c(p_f)\gamma^0\ks\as_2]\}B_{1s}B_{2s}^*.
\label{42}
\end{eqnarray}
From now on, we define a 4-vector
\begin{equation}
\eta^{\mu}=(1,0,0,0). \label{43}
\end{equation}
We can therefore write
\begin{equation}
\gamma^0=\ets. \label{44}
\end{equation}
Then, Eq.(\ref{42}) becomes
\begin{eqnarray}
\mathcal{M}_6&=&Tr\{(\ps_f c+c^2)C_1(\ps_i
c+c^2)\oll{C}_2\}B_{1s}B_{2s}^*\nonumber\\
&=&Tr\{(\ps_f c+c^2)[c(p_i)\ets\ks\as_1+c(p_f)\as_1\ks\ets](\ps_i
c+c^2)\nonumber\\
&&[c(p_i)\as_2\ks\ets+c(p_f)\ets\ks\as_2]\}B_{1s}B_{2s}^*.
\label{45}
\end{eqnarray}
In \cite{1}, the authors claim that the controversial
$\sin(2\phi_0)$ term disappear because it is proportional to terms
like $Tr\{(\ps_f c+c^2)\gamma^0\ks\as_1(\ps_i
c+c^2)\as_2\ks\gamma^0\}$. This term as well as $Tr\{(\ps_f
c+c^2)\as_1\ks\gamma^0(\ps_i c+c^2)\gamma^0\ks\as_2\}$ are indeed
zero but for $Tr\{(\ps_f c+c^2)\gamma^0\ks\as_1(\ps_i
c+c^2)\gamma^0\ks\as_2\}$ and $Tr\{(\ps_f
c+c^2)\as_1\ks\gamma^0(\ps_i c+c^2)\as_2\ks\gamma^0\}$ this is no
longer true. These terms are not zero and we give explicitly
their values
\begin{eqnarray}
Tr\{(\ps_f c+c^2)\gamma^0\ks\as_1(\ps_i
c+c^2)\gamma^0\ks\as_2\}&=&Tr\{(\ps_f c+c^2)\as_1\ks\gamma^0(\ps_i
c+c^2)\as_2\ks\gamma^0\}\nonumber\\
&=&8w^2\{(a_1.p_f)(a_2.p_i)+(a_1.p_i)(a_2.p_f)\}. \label{46}
\end{eqnarray}
In most cases, the various traces are zero except when the cyclic
process of taking scalar products of pairs comes to products such
that
\begin{eqnarray}\left.\begin{array}{c}
(k.\eta)(k.\eta)(a_1.p_i)(a_2.p_f),\\
(k.\eta)(k.\eta)(a_1.p_f)(a_2.p_i), \label{47}
\end{array}\right.
\end{eqnarray}
in which case, one has contributions proportional to
$w^2(a_1.p_i)(a_2.p_f)$ and $w^2(a_1.p_f)(a_2.p_i)$ respectively.
Explicitly, we give the result for $\mathcal{M}_6$ and
$\mathcal{M}_8$. \\One has
\begin{eqnarray}
\mathcal{M}_6&=&\frac{w^2}{c^2}\{2\sin(2\phi_0)\left[\frac{(a_1.p_i)}{(k.p_i)}\frac{(a_2.p_f)}{(k.p_f)}+\frac{(a_2.p_i)}{(k.p_i)}\frac{(a_1.p_f)}{(k.p_f)}\right]J_{s+1}J_{s-1}\nonumber\\
&&+i[-\{(a_1.p_i)(a_2.p_f)+(a_1.p_f)(a_2.p_i)\}J^2_{s-1}\nonumber\\
&&+\{(a_1.p_i)(a_2.p_f)+(a_1.p_f)(a_2.p_i)\}J^2_{s+1}]\},
\label{48}
\end{eqnarray}
while $\mathcal{M}_8$ is given by
\begin{eqnarray}
\mathcal{M}_8&=&\frac{w^2}{c^2}\{2\sin(2\phi_0)\left[\frac{(a_1.p_i)}{(k.p_i)}\frac{(a_2.p_f)}{(k.p_f)}+\frac{(a_2.p_i)}{(k.p_i)}\frac{(a_1.p_f)}{(k.p_f)}\right]J_{s+1}J_{s-1}\nonumber\\
&&-i[-\{(a_1.p_i)(a_2.p_f)+(a_1.p_f)(a_2.p_i)\}J^2_{s-1}\nonumber\\
&&+\{(a_1.p_i)(a_2.p_f)+(a_1.p_f)(a_2.p_i)\}J^2_{s+1}]\}.
\label{49}
\end{eqnarray}
The fact that complex numbers appear in the expressions of
$\mathcal{M}_6$ and $\mathcal{M}_8$ is not surprising since the
former is the complex conjugate of the latter and their real sum
is such that
\begin{eqnarray}
\mathcal{M}_6+\mathcal{M}_8&=&\frac{4w^2}{c^2}\sin(2\phi_0)\left[\frac{(a_1.p_i)}{(k.p_i)}\frac{(a_2.p_f)}{(k.p_f)}+\frac{(a_2.p_i)}{(k.p_i)}\frac{(a_1.p_f)}{(k.p_f)}\right]J_{s+1}J_{s-1}.
\label{50}
\end{eqnarray}

So, the first controversy is settled and there is indeed a term
containing $\sin(2\phi_0)$ in the expression of the scattering
cross section.  We have written a Reduce program that calculates
analytically the traces in Eq.(\ref{37}). Before writing our
Reduce program, we have extensively studied the textbook by A. G.
Grozin \cite{5} which is full of worked examples in various
fields of physics particularly in QED. We give the final result
for the unpolarized DCS for the Mott scattering of a Dirac-Volkov
electron :\\
\begin{eqnarray}
\frac{d\oll{\sigma}^{(s)}}{d\Omega_f}&=&\frac{Z^2}{c^2}\frac{|\mathbf{q}_f|}{|\mathbf{q}_i|}\frac{1}{|\mathbf{q}_f-\mathbf{q}_i-s\mathbf{k}|^4}\nonumber\\
&&\times \frac{2}{c^2}\left\{J_s^2A+
\big(J^2_{s+1}+J^2_{s-1}\big)B
+\big(J_{s+1}J_{s-1}\big)C+J_s\big(J_{s-1}+J_{s+1}\big)D\right\},
\label{51}
\end{eqnarray}
where for notational simplicity we have dropped the argument $z$
in the various ordinary Bessel functions. The coefficients $A$,
$B$, $C$ and $D$ are respectively given by
\begin{eqnarray}
A&=&c^4-(q_f.q_i)c^2+2Q_fQ_i-\frac{a^2}{2}\left(\frac{(k.q_f)}{(k.q_i)}+\frac{(k.q_i)}{(k.q_f)}\right)+\frac{a^2\omega^2}{c^2(k.q_f)(k.q_i)}((q_f.q_i)-c^2)+\nonumber\\
&&\frac{(a^2)^2\omega^2}{c^4(k.q_f)(k.q_i)}+\frac{a^2\omega}{c^2}(Q_f-Q_i)\left(\frac{1}{(k.q_i)}-\frac{1}{(k.q_f)}\right),
\label{52}
\end{eqnarray}
\begin{eqnarray}
B&=&-\frac{(a^2)^2\omega^2}{2c^4(k.q_f)(k.q_i)}+\frac{\omega^2}{2c^2}\left(\frac{(a_1.q_f)}{(k.q_f)}\frac{(a_1.q_i)}{(k.q_i)}+\frac{(a_2.q_f)}{(k.q_f)}\frac{(a_2.q_i)}{(k.q_i)}\right)-\frac{a^2}{2}+\nonumber\\
&&\frac{a^2}{4}(\frac{(k.q_f)}{(k.q_i)}+\frac{(k.q_i)}{(k.q_f)})-\frac{a^2\omega^2}{2c^2(k.q_f)(k.q_i)}\big((q_f.q_i)-c^2\big)+\nonumber\\
&&\frac{a^2\omega}{2c^2}(Q_f-Q_i)\left(\frac{1}{(k.q_f)}-\frac{1}{(k.q_i)}\right),
\label{53}
\end{eqnarray}
\begin{eqnarray}
C&=&\frac{\omega^2}{c^2(k.q_f)(k.q_i)}\big(\cos(2\phi_0)\{(a_1.q_f)(a_1.q_i)-(a_2.q_f)(a_2.q_i)\}+\nonumber\\
&&\sin(2\phi_0)\{(a_1.q_f)(a_2.q_i)+(a_1.q_i)(a_2.q_f)\}\big),
\label{54}
\end{eqnarray}
\begin{eqnarray}
D&=&\frac{c}{2}\left((\AA.q_i)+(\AA.q_f)\right)-\frac{c}{2}\left(\frac{(k.q_f)}{(k.q_i)}(\AA.q_i)+\frac{(k.q_i)}{(k.q_f)}(\AA.q_f)\right)+\nonumber\\
&&\frac{\omega}{c}\left(\frac{Q_i(\AA.q_f)}{(k.q_f)}+\frac{Q_f(\AA.q_i)}{(k.q_i)}\right),
\label{55}
\end{eqnarray}
where $\AA=a_1\cos(\phi_0)+a_2\sin(\phi_0)$.
\subsection{Comparison of the coefficients.}
The argument about the missing term proportional to
$\sin(2\phi_0)$ having been given a convincing explanation, we now
turn to other remarks along the same lines since there are indeed
other differences between our result and the result of \cite{1}.
We discuss now the difference occurring in our expression of the
coefficient $A$ and the corresponding one of \cite{1}. In their
expression multiplying the product $2J_n^2(\xi)$, the single term
$\frac{(a^2)^2w^2}{c^6(k.q)(k.q')}$ should come with a coefficient
$\frac{1}{2}$. We have written a second Reduce program that allows
 the comparison between the coefficient $A$ of \cite{1} and
the coefficient $A$ of this work. There are so many differences
between our result and the result they found for the coefficient
$B$ that we refer the reader to our main Reduce program \cite{8}.
The coefficient $C$ has already been discussed. As for the
coefficient $D$, we have found an expression that is linear in
the electromagnetic potential. In a third Reduce program, it is
shown explicitly that if we ignore the first term in the
coefficient multiplying $J_s(J_{s-1}+J_{s+1})$ given in \cite{1},
one easily gets the result we have obtained. This term does not
come from the passage from the variables $(p,\tilde{p})$ to the
variable $(q,\tilde{q})$. The introduction of such 4-vector
$\tilde{q}$ is not useful, makes the calculations rather lengthy
and gives rise to complicated expressions. As a supplementary
consistency check of our procedure used in writing the main
Reduce program, we have reproduced the result of the DCS
corresponding to the Compton scattering in an intense
electromagnetic field given by Berestetzkii, Lifshitz and
Pitaevskii \cite{6}.
\section{Results and discussion}
\subsection{Kinematics of the collision}

 For the description of the scattering geometry, we work in
a coordinate system in which $\mathbf{k}||\widehat{e}_z$. This
means that the direction of the laser propagation is along the
$Oz$ axis. To avoid any confusion, we will compare the
Dirac-Volkov DCS (26) of \cite{1} with the corresponding DCS
(\ref{34}) we have obtained in the same coordinate system. The
spinless DCS will also be discussed as well as the non
relativistic one. We begin by defining our scattering geometry.
In our system, the vector $\textbf{p}_i$ is such that
$\textbf{p}_i||\widehat{e}_x$, meaning that the undressed angular
coordinate of the incoming electron are
$\theta_i=90^{\circ},\,\phi_i=0^{\circ}$. For the scattered
electron, the vector $\textbf{p}_f$ is such $\textbf{p}_f \in
(\widehat{e}_y,\widehat{e}_z)$, meaning that $-180^{\circ}\leq
\theta_f\leq 180^{\circ}$ and $\phi_f=90^{\circ}$. With this
choice, we have been able to reproduce qualitatively the results
and all the figures of \cite{1}. The reason underlying this
choice of the coordinates is the following. The angles
$(\theta_i,\phi_i)$ of $\textbf{p}_i$ (the same holds for
$\textbf{p}_f$) are the intrinsic angular coordinates of the
incoming electron. As the vector $\textbf{q}_i$ is defined
through $\textbf{p}_i$ via the relation
\begin{equation}
\mathbf{q}_i=\mathbf{p}_i-\frac{a^2}{2(k.p_i)c^2}\mathbf{k},
\label{56}
\end{equation}
we cannot define intrinsic angular coordinates using
$\mathbf{q}_i$. When the electron is subjected to the radiation
field, it acquires new angular coordinates that can easily be
determined. The key quantity that gives an idea of the dependence
of $\mathbf{q}_i$(and $\mathbf{q}_f$) on the spatial orientation
of the electron momentum due to ($k.p_i$)(and ($k.p_f$)) is the
cosine of the angle between $\mathbf{q}_i$ and $\mathbf{q}_f$.
While
\begin{equation}
\mathbf{p}_i.\mathbf{p}_f=|\mathbf{p}_i||\mathbf{p}_f|\cos(\widehat{\mathbf{p}_i,\mathbf{p}_f}),
\label{57}
\end{equation}
with
\begin{equation}
cos(\widehat{\mathbf{p}_i,\mathbf{p}_f})=sin(\theta_{i})sin(\theta_{f})cos(\phi_{i}-\phi_{f})+cos(\theta_{i})cos(\theta_{f}),
\label{58}
\end{equation}
we have
\begin{equation}
\mathbf{q}_i.\mathbf{q}_f=|\mathbf{q}_i||\mathbf{q}_f|\cos(\widehat{\mathbf{q}_i,\mathbf{q}_f}),
\label{59}
\end{equation}
with
\begin{eqnarray}
|\mathbf{q}_i||\mathbf{q}_f|\cos(\widehat{\mathbf{q}_i,\mathbf{q}_f})
&=&|\mathbf{p}_i||\mathbf{p}_f|\cos(\widehat{\mathbf{p}_i,\mathbf{p}_f})
-\frac{a^{2}w}{2c^{3}}\left(\frac{|\mathbf{p}_{i}|\cos(\theta_{i})}{(k.p_{f})}+\frac{|\mathbf{p}_{f}|\cos(\theta_{f})}{(k.p_{i})}\right)\nonumber\\
&+&\frac{(a^2)^2w^2}{4(k.p_f)(k.p_i)c^6}. \label{60}
\end{eqnarray}
From these relations, one deduce that in the limit of low
incoming electron energies and moderate field strength,
$\cos(\widehat{\mathbf{q}_i,\mathbf{q_f}})$ and
$\cos(\widehat{\mathbf{p}_i,\mathbf{p_f}})$ are very close
therfore the second and third term of the RHS of Eq.(\ref{60})
can be safely neglected. For high incoming electron energies and
intense field strength, the difference between the two cosines
increases and these terms cannot be neglected. We shall give for
the sake of illustration, tables that compare these two cosines
for the three regimes we shall investigate, namely the non
relativistic-moderate field strength regime, the
relativistic-strong field strength regime and finally the
relativistic-intense field strength regime. We choose the same
value as \cite{1} for the laser angular frequency $w=0.0430 \,a.u$
for all the numerical calculations. This typical near infra-red
angular frequency is that of a neodymium laser. The other
parameters are the electric field strength $\varepsilon$ and the
relativistic parameter
$\gamma=1/\sqrt{1-\beta^2}=1/\sqrt{1-v^2/c^2}$. This parameter
fixes the incoming electron total energy $E_i$ via the relation
$E_i=c^2\gamma$ from which one deduces the corresponding incoming
electron kinetic $T_i$ by subtracting the rest energy $(c^2$ in
a.u) $T_i=E_i-c^2=c^2(\gamma-1)$. Before beginning our
discussion, we would like to make general comments on the figures
obtained in \cite{1} starting with Figure  \ref{figure3}. This
figure does not represent the envelope of the controversial
generalized equation (26) of that work. Indeed, we shall see that
it represents the envelope of the non relativistic DCS given by
Eq.(34) in \cite{1}. We give the correct envelope for the
relativistic calculations obtained by using either Eq.(\ref{34})
of our work or Eq.(26) of \cite{1}. In Figure (\ref{6}) of
\cite{1}, there is a difference between the Dirac-Volkov DCS (26)
and the spinless particle DCS (30) though the overall behaviour is
smoothly oscillatory. The results we have obtained show the same
oscillatory behaviour. The curves for the Dirac-Volkov DCS (26)
of \cite{1} and the Dirac-Volkov DCS (\ref{34}) of our work are
almost identical while the difference between two relativistic
DCSs and the spinless particle DCS given by Eq.(30) of \cite{1}
is less important than in Figure 6 of \cite{1}. Figure  7 of
\cite{1} is the only figure we agree with. In Figure 8 of
\cite{1}, we disagree with the behaviour of the Dirac-Volkov DCS
(26) of \cite{1} particularly for small angles around
$\theta_f=0^{\circ}$. When programing Eq.(26) of \cite{1}, we
obtained a value for the Dirac-Volkov DCS at $\theta_f=0^{\circ}$
of nearly $3.2\,10^{-14}$ $a.u$ instead of the $2.2\,10^{-14}$
$a.u$ indicated in Figure 8 of \cite{1}. Moreover, the electric
field strength $\varepsilon$ being a key parameter (as well as the
incoming electron total energy), we have compared our
Dirac-Volkov DCS and the Dirac-Volkov DCS (26) of \cite{1} and we
have come to the following important conclusions. First, for the
non relativistic and low and low-intensity field strength regime
($\gamma=1.0053\,a.u,\,\varepsilon=0.05\,a.u$) and for the
relativistic regime and increasing field strength
($\gamma=2.00\,a.u,\,\varepsilon=1.00\,a.u$) the differences
between our results and the results found in \cite{1} are small
but approach one percent. Second, we have a different picture for
the relativistic-high intensity regime
($\gamma=2.00\,a.u,\,\varepsilon=5.89\,a.u$) where the missing
terms in \cite{1} lead to values of the Dirac-Volkov DCS (26) of
\cite{1} that over-estimate the corresponding DCS (\ref{34})of
our work. Even in the non relativistic regime ($\gamma=1.0053\,
a.u$) but for increasing field strength, the difference between
our results and the results of \cite{1} begins to appear
clearly.\\
We turn now to a qualitative and quantitative discussion of the
physical process. We shall comment and analyze the results
obtained in \cite{1} in the light of those we have obtained
bearing in mind that we can hardly escape rephrasing the physical
insights and explanations contained in \cite{1}. Our disagreement
is quantitative since we have shown in the first part of this
work that the expression (26) of \cite{1} contains errors and a
missing term proportional to $\sin(2\phi_0)$. So, our primary
task is to assess the importance of this errors and missing term
and to what extent they modify the quantitative and qualitative
contents of \cite{1}.
\subsection{The non relativistic-low electric field strength regime}
\begin{figure}[ht]
\centering
\includegraphics[angle=0,width=3 in,height=2.3 in]{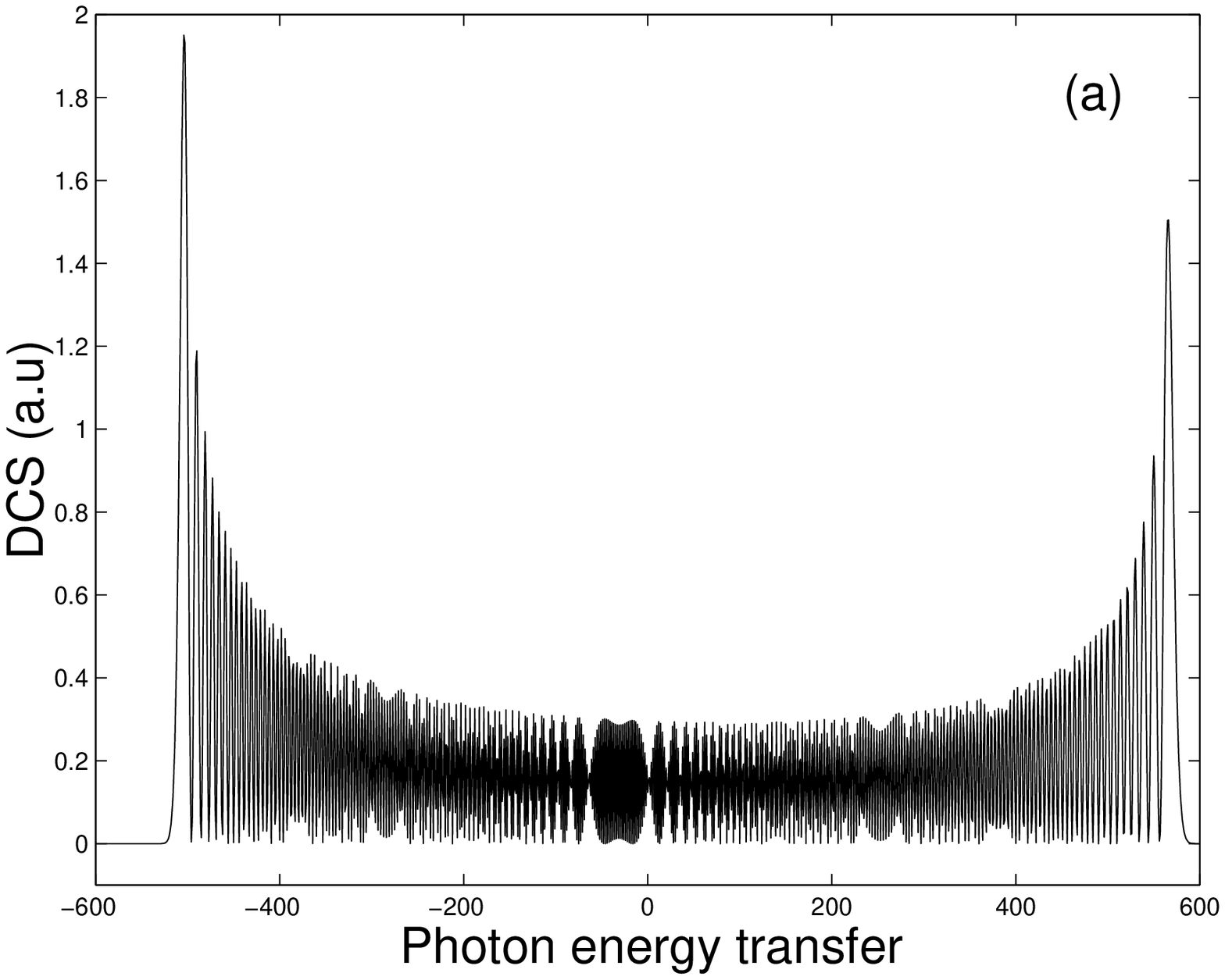}
\includegraphics[angle=0,width=3 in,height=2.3 in]{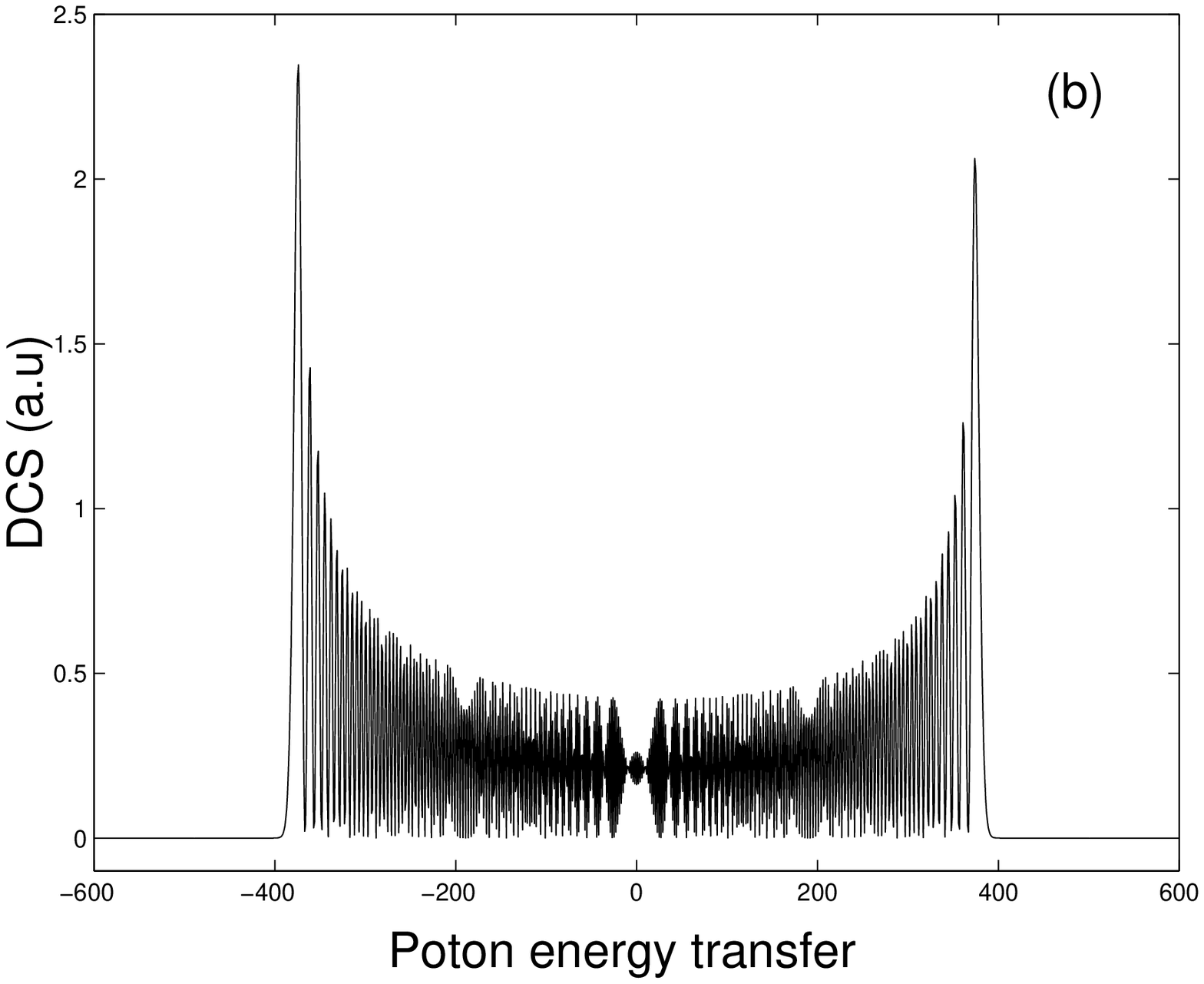}
\caption{\label{figure1} (a): Envelope of the non relativistic
differential cross-section $d\sigma/d\Omega$ scaled in unit of
$10^{-7}$ $a.u$ as a function of energy transfer $Q_f-Q_i$ scaled
in units of the laser photon energy $w$ for an electric field
strength of $0.05\,a.u$. and a relativistic parameter
$\gamma=1.0053\,a.u$, (b). Envelope of the relativistic
differential cross-section $(d\sigma/d\Omega)_{DV}$ scaled in unit
of $10^{-7}$ $a.u$ as a function of the laser photon energy $w$
for the same parameters. The envelope for
$(d\sigma/d\Omega)_{DV}^{\cite{1}}$ and $(d\sigma/d\Omega)_{KG}$
are almost identical}
\end{figure}

In this regime, we choose as in \cite{1} $\gamma=1.0053\,a.u$ for
the relativistic parameter and $\varepsilon=0.05\,a.u$ for the
electric field strength. This relativistic parameter corresponds
to an incoming electron kinetic energy $T_i=100\,a.u=2.700\,kev$.
With our choice of the angular parameters, we compare in Table.
\ref{tabel1} some values of
$\cos(\widehat{\mathbf{p}_i,\mathbf{p}_f})$ and
$\cos(\widehat{\mathbf{q}_i,\mathbf{q}_f})$.
\begin{center}
\begin{tabular}{|c|c|}
\hline\hline
$\cos(\widehat{\mathbf{p}_i,\mathbf{p}_f})$&$\cos(\widehat{\mathbf{q}_i,\mathbf{q}_f})$\\
\hline\hline
0.853553&0.853631\\
\hline\hline
0.850868&0.850942\\
\hline\hline
0.847923&0.847993\\
\hline\hline
0.844719&0.844786\\
\hline\hline
 0.841259&0.841322\\
 \hline\hline
\end{tabular}
\begin{table}[h]
\caption{\label{tabel1}}
\end{table}
\end{center}
\indent The difference is small so in this regime and the
coordinates of the undressed electron are nearly the same as that
of the dressed electron. We plot in the upper part (a) of Figure
\ref{figure1} the non relativistic DCS given by Eq.(34) of
\cite{1} and in the lower part (b) of the same figure, the
generalized Dirac-Volkov DCS given either by Eq.(26) of \cite{1}
or Eq.(\ref{34}) of our work as a function of the final electron
energy scaled to the photon energy. The scattering angle is large
enough so that an important number of photons can be exchanged in
the course of the collision. In this low-intensity regime, the
envelope of the non relativistic DCS is qualitatively different
from the envelope for the Dirac-Volkov and Klein-Gordon DCSs.

 The observed cutoffs occur at $s_{min}=-522$ and
$s_{max}=582$ for the non relativistic DCS and $s_{min}=-474$ and
$s_{max}=474$ both for the Dirac-Volkov and Klein-Gordon DCSs
since the argument that appears in the ordinary Bessel functions
is the same for both DCSs. So the comments made in \cite{1}
concerning the interpretation of the envelope obtained do not
apply for the Dirac-Volkov and Klein-Gordon cases. While the
spectrum of Figure (1.a) of our work (which is identical to that
of Figure (1.a) of \cite{1}) exhibits an overall asymmetric
envelope with peaks of negative energy transfer higher than peaks
of positive energy transfer, this asymmetry is less pronounced in
the case of the Dirac-Volkov and spinless particle DCSs. This
emphasized asymmetry in the non relativistic case can easily be
traced back by a close look at Eq.(34) of \cite{1}. Indeed, the
non relativistic DCS depends on $J_s^2(z)$ ($z$ depends only
weakly on $s$) so the asymmetry can only come from the dependence
of the modulus of the final momentum $\mathbf{q}_f$ on the number
of the transferred photons $s$ according to Eq.(25) of \cite{1}.
We explicitly write this equation (with our notation)
\begin{equation}
|\mathbf{q}_f|=\left(|\mathbf{q}_i|^2+2s\frac{wQ_i}{c^2}+s^2\frac{w^2}{c^2}\right)^{1/2}.
\label{61}
\end{equation}
Also, the denominator
$|\mathbf{q}_i-\mathbf{q}_f+s\mathbf{k}|^{4}$ depends linearly on
the number $s$ of transferred photons and this gives rise to an
asymmetric envelope for the non relativistic DCS. In Figure 2, we
plot the behaviour of
$|\mathbf{q}_i-\mathbf{q}_f+s\mathbf{k}|^{-4} $ as a function of
the number $s$ of photons transferred.
\begin{figure}[ht]
\centering
\includegraphics[angle=0,width=3 in,height=2.3 in]{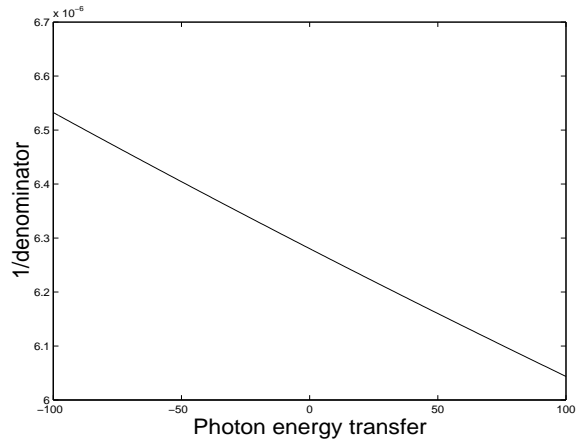}
\caption{ \label{figure2}Behaviour of
$|\mathbf{q}_i-\mathbf{q}_f+s\mathbf{k}|^{-4} $ as a function of
the number $s$ of the photons transferred with $\gamma=1.0053$ and
$\varepsilon=0.05\, a.u$}
\end{figure}
\indent As mentioned in \cite {1}, we have an enhancement of
negative over positive-energy transfer cross-section. The DCSs
fall of abruptly beyond the points where the argument of the
Bessel functions equal to the order. For the Dirac-Volkov and
Klein-Gordon DCSs, this cutoff occurs ( up to machine precision)
numerically for $s=\pm 474$ and an argument $z$ of the ordinary
Bessel functions almost constant and equal to $380.016$. However
Figure (1.b) shows a visual cutoff for $s=\pm 392$ since the
infinitesimal contributions to the DCSs cannot be plotted. For
the non relativistic DCS, the numerical cutoff occurs (again up
to machine precision) for $s=-606$ and $s= 685$. The visual cutoff
occurs for $s=-500$ and $s= 590$. The difference  $\Delta s=20$
between our results and that of \cite {1} is just a matter of
convention. We now analyze the angular distributions. We have
summed as in \cite{1} $\pm 100$ peaks around the elastic one in
order to draw the angular dependence of the DCS. In Figure 6. of
\cite{1}, the accumulated DCS is shown for an electric field
strength $\varepsilon=0.05\,a.u$. The computer code we have
written calculates the Dirac-Volkov DCS (\ref{34}) of our  work,
the Dirac-Volkov DCS (26) of \cite{1}, the spinless particle DCS
and the non relativistic DCS. At least, in the non relativistic
regime, our results and that of \cite{1} agree very well and are
both close to the results for a spinless particle. We give in
Figure  \ref{figure3} the angular distribution of the various
DCSs.

\begin{figure}[ht]
\centering
\includegraphics[angle=0,width=3 in,height=2.3 in]{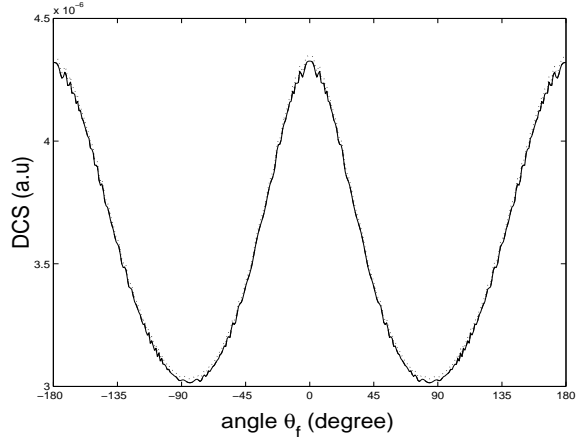}
\caption{ \label{figure3} Summed differential cross sections
$d\sigma/d\Omega$ of $\pm 100$ peaks around the elastic one as a
function of the angle $\theta_f$ for a relativistic parameter
$\gamma=1.0053\, a.u$ and an electric field strength
$\varepsilon=0.05\, a.u$. The solid line denotes the result for
Dirac-Volkov electrons, the long dashed one sketches the values
for $(d\sigma/d\Omega)_{DV}^{[1]}$ and the short dashed is the
result for spinless particles }
\end{figure}
\indent Apart from minor differences, all three calculations
exhibit maxima for $\theta_f=0^{\circ} $ and $\pm 180^{\circ}$, a
giggling oscillatory behaviour (as in \cite{1}) and minima
slightly shifted from $\pm 90^{\circ}$ ( at $ \pm 84^{\circ}$).
Let aside the order of magnitude, we have in our case, three DCSs
that  are close to each other and not as differentiated as shown
in Figure 6. of \cite{1}. We do not agree at all with the results
shown in Figure 6. In particular, our extrema for the various
DCSs are shown in table (\ref{table2}) (scaled in $10^{-6}\,
a.u$).
\begin{center}
\begin{tabular}{|c|c|c|c|}
\hline\hline
$\theta_f$&$(d\sigma/d\Omega)_{DV}$&$(d\sigma/d\Omega)_{DV}^{[1]}$&$(d\sigma/d\Omega)_{KG}$\\
\hline\hline
$-180^{\circ}$&4.32027&4.32027&4.34311\\
\hline\hline
-$84^{\circ}$&3.01486&3.01486&3.03079\\
\hline\hline
$0^{\circ}$&4.326&4.326&4.34886\\
\hline\hline
$84^{\circ}$&3.01486&3.01486&3.03079\\
\hline\hline
 $180^{\circ}$&4.32027&4.32027&4.34311\\
 \hline\hline
\end{tabular}
\begin{table}[h]
\caption{\label{table2}}
\end{table}
\end{center}
\indent So, this adds to the controversy. Even if we use the
expression for the Dirac-Volkov DCS given by Eq.(26) of \cite{1},
we have a different figure for the non relativistic regime. If we
now increase the electric field strength from $\varepsilon=
0.05\, a.u$ to  $ \varepsilon= 1.00\, a.u$, the agreement remains
good between the three relativistic calculations. There is still
a maximum at $\theta_f=0^{\circ} $ while the minima are shifted
towards $\pm 117^{\circ}$. To give an idea the small differences
between our result and the result of \cite{1} for the
Dirac-Volkov DCS, we have plotted in the upper part (a) of Figure
4, the ratio of the DCS given by Eq.(26) of \cite{1} to the DCS
given by Eq.(\ref{34}) of our work as a function of the angle
$\theta_f$ for $ \varepsilon= 1.00\,u.a$. The ratio $R$ is
defined by
\begin{equation}
R=\frac{(\frac{d\sigma}{d\Omega_{f}})_{DV}^{[1]}}{(\frac{d\sigma}{d\Omega_{f}})_{DV}}.
\label{62}
\end{equation}
\begin{figure}[ht]
\centering
\includegraphics[angle=0,width=3 in,height=2.4 in]{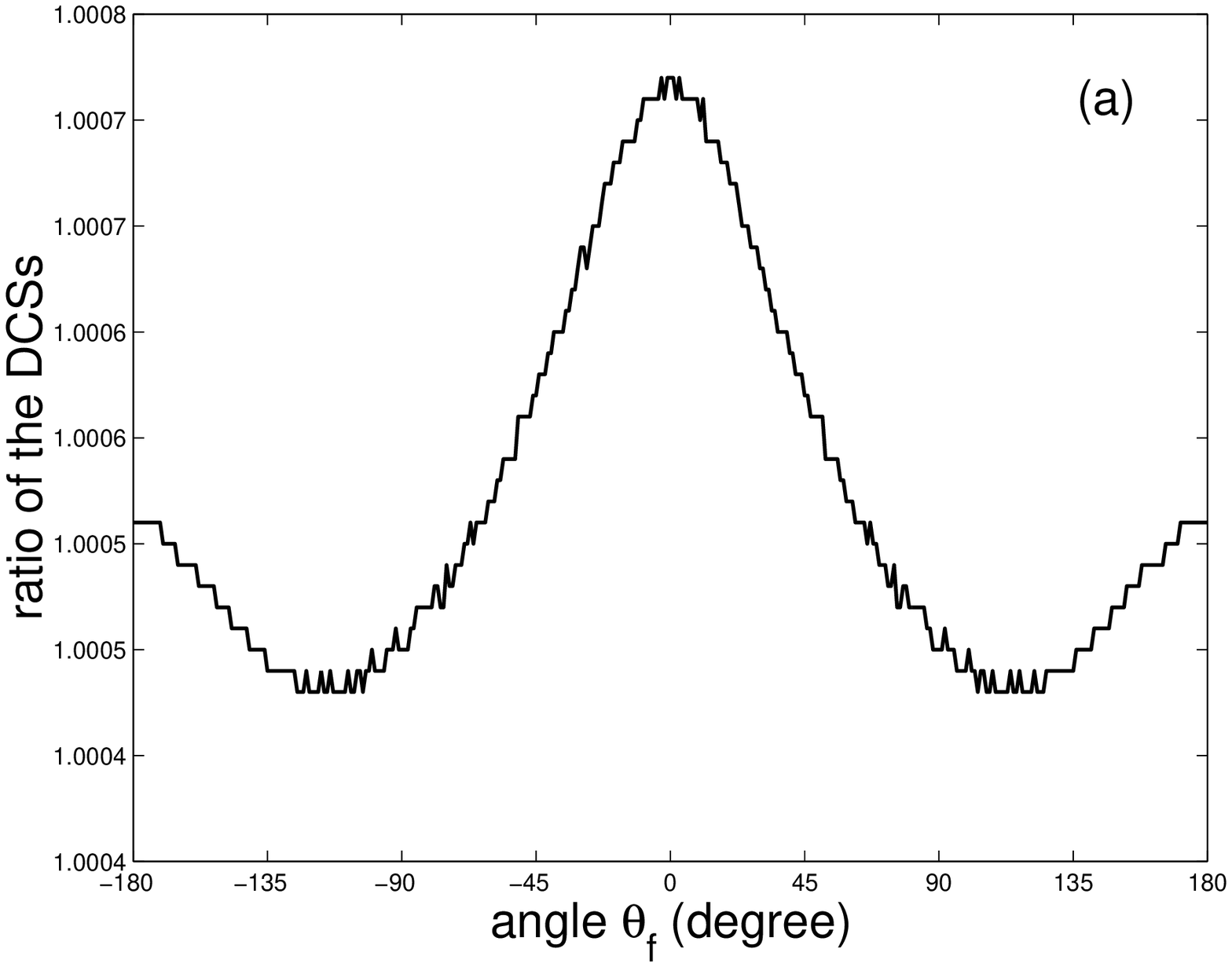}
\includegraphics[angle=0,width=3 in,height=2.4 in]{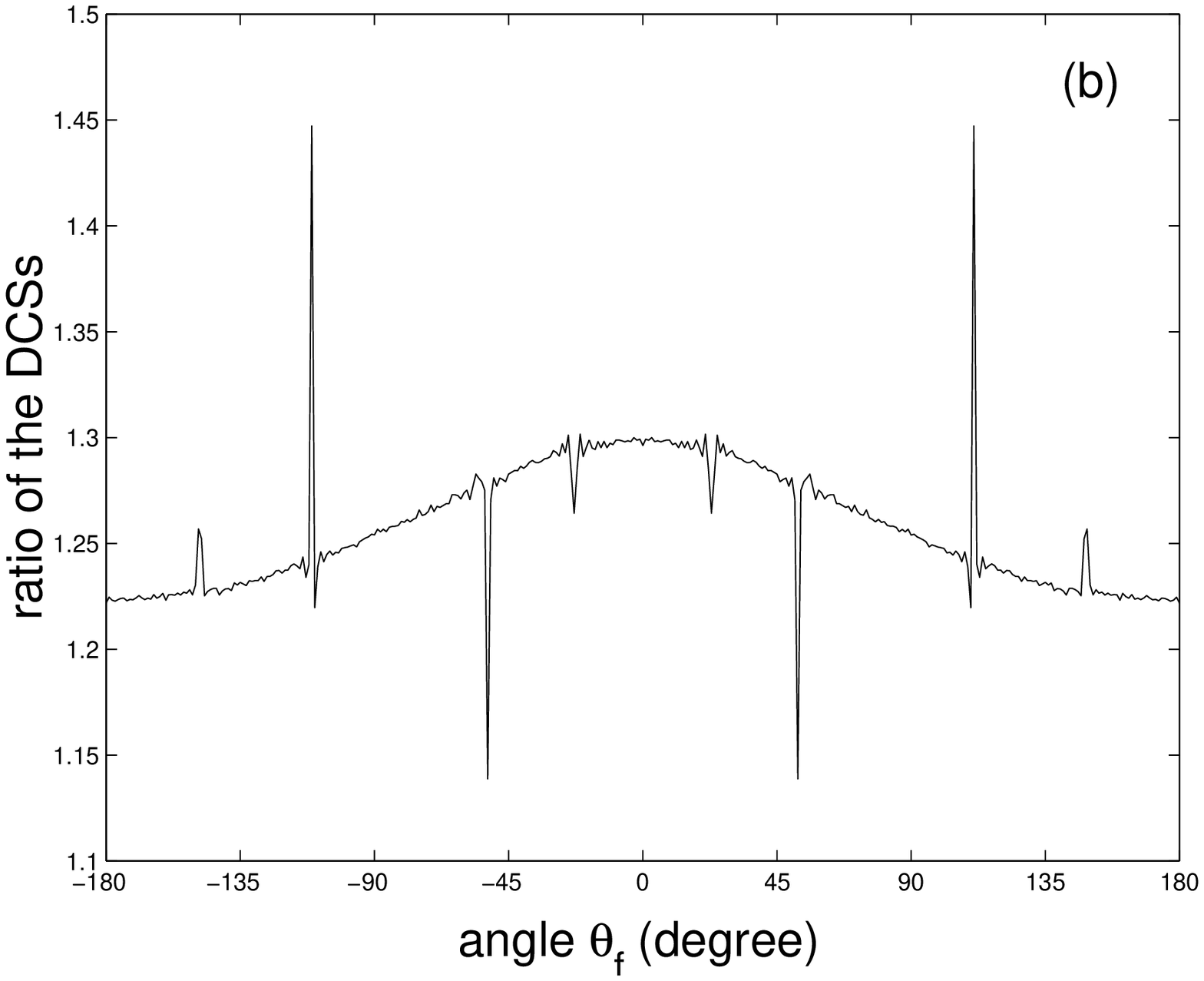}
\caption{(a) Ratio $R$ of the two Dirac-Volkov DCS for
$\gamma=1.0053,\,\varepsilon=1.00\,a.u$ and $s=\pm 100$. (b):
Ratio $R$ of the two Dirac-Volkov DCS for $\gamma=1.0053\,a.u
,\,\varepsilon=5.00\, a.u$ and $s=\pm 100$}
\end{figure}
The deviations from the expected value 1 are shown and have the
same shape as the corresponding DCS. However, for increasing
electric field strength, the values for this ratio are not close
to 1. For a relativistic parameter $\gamma=1.0053\,a.u$ and for
an electric field strength $ \varepsilon= 5.00\, u.a$ and $s=\pm
 100$, our results for the Dirac-Volkov DCS and the corresponding
results of  \cite{1} do not agree at all. In the lower part (b)
of Figure (4), there is an over estimation varying from $22.5\%$
to $30\%$ with some peaks giving an over estimation of up to
$45\%$ for the DCS (26) of \cite{1} compared to the corresponding
DCS (\ref{34}) of this work. All these peaks are nearly multiples
or submultiples of an angle close to $\pi/4$.
\subsection{Relativistic-strong electric field strength regime}
 For the relativistic regime, we have chosen the parameters of
\cite{1} $\gamma=2$ which corresponds to an incoming electron
total energy $E_i=2c^2$ or a $T_i=0.5116\,MeV$. The electric field
strength is now $\varepsilon=1.00\,a.u$. Some cosines of the
angles $\cos(\widehat{\mathbf{q}_i,\mathbf{q}_f})$ and
$\cos(\widehat{\mathbf{p}_i,\mathbf{p}_f})$ are shown in Table
(\ref{table3})
\begin{center}
\begin{tabular}{|c|c|}
\hline
$\cos(\widehat{\mathbf{p}_i,\mathbf{p}_f})$&$\cos(\widehat{\mathbf{q}_i,\mathbf{q}_f})$\\
\hline\hline
0.853553&0.855757\\
\hline\hline
0.850868&0.852775\\
\hline\hline
0.847923&0.849543\\
\hline\hline
0.844719&0.846062\\
\hline\hline
 0.841259&0.842332\\
\hline\hline
\end{tabular}
\begin{table}[h]
\caption{\label{table3}}
\end{table}
\end{center}
 \indent In this regime, dressing effects are important. The
envelope of the energy distribution of the scattered electrons is
similar to the one displayed in the lower part (b) of Figure
$\ref{figure1}$. However, there is a more important asymmetry
than in the non relativistic regime with was to be expected. The
corresponding cutoffs are $(-170000, 194000)$ for
${(\frac{d\sigma}{d\Omega_{f}})}_{NR}$ and $(-66000, 64000)$ for
${(\frac{d\sigma}{d\Omega_{f}})}_{DV}$,${(\frac{d\sigma}{d\Omega_{f}})}_{DV}^{[1]}$
and ${(\frac{d\sigma}{d\Omega_{f}})}_{KG}$. The three
relativistic calculations lead to angular distributions peaked in
the direction of the laser propagation $\theta_{f}= 0^{\circ}$.
The two Dirac-Volkov DCSs (solid line and long dashed line) are
slightly different only in the vicinity of the two minima located
at $\theta_{f}\simeq \pm 33^{\circ}$. In this regime, the shape of
the ratio $R$ is similar to that of the corresponding DCSs. This
ratio is equal to $1$ for $\theta_{f}= 180^{\circ}$ but there is
now an overall amplitude of $8. 10^{-3}$ around the expected value
$1$. If we increase the electric field strength from $\varepsilon
= 1.00 \,a.u $ to $\varepsilon= 5.89 \,a.u$ and keep the same
value of the relativistic parameter $\gamma=2$, the difference
between our Dirac-Volkov results and the corresponding results of
$[1]$ becomes important.\\
 In the upper part(a) of Figure 5, we give
 the two Dirac-Volkov DCSs and in the lower part (b) of the
same figure we give the ratio $R$ of the two DCSs for the
relativistic parameter $\gamma=2.00$ and an electric field
strength $\varepsilon =5.89 \,a.u$. For the angles $\theta_{f}=\pm
180^{\circ}$, the ratio is $R\simeq 1.01$ while for the peak in
the direction of the laser propagation, $\theta_{f}=0^{\circ}$,
the ratio is $R\simeq 2.34$.
\begin{figure}[ht]
\centering
\includegraphics[angle=0,width=3 in,height=2.3 in]{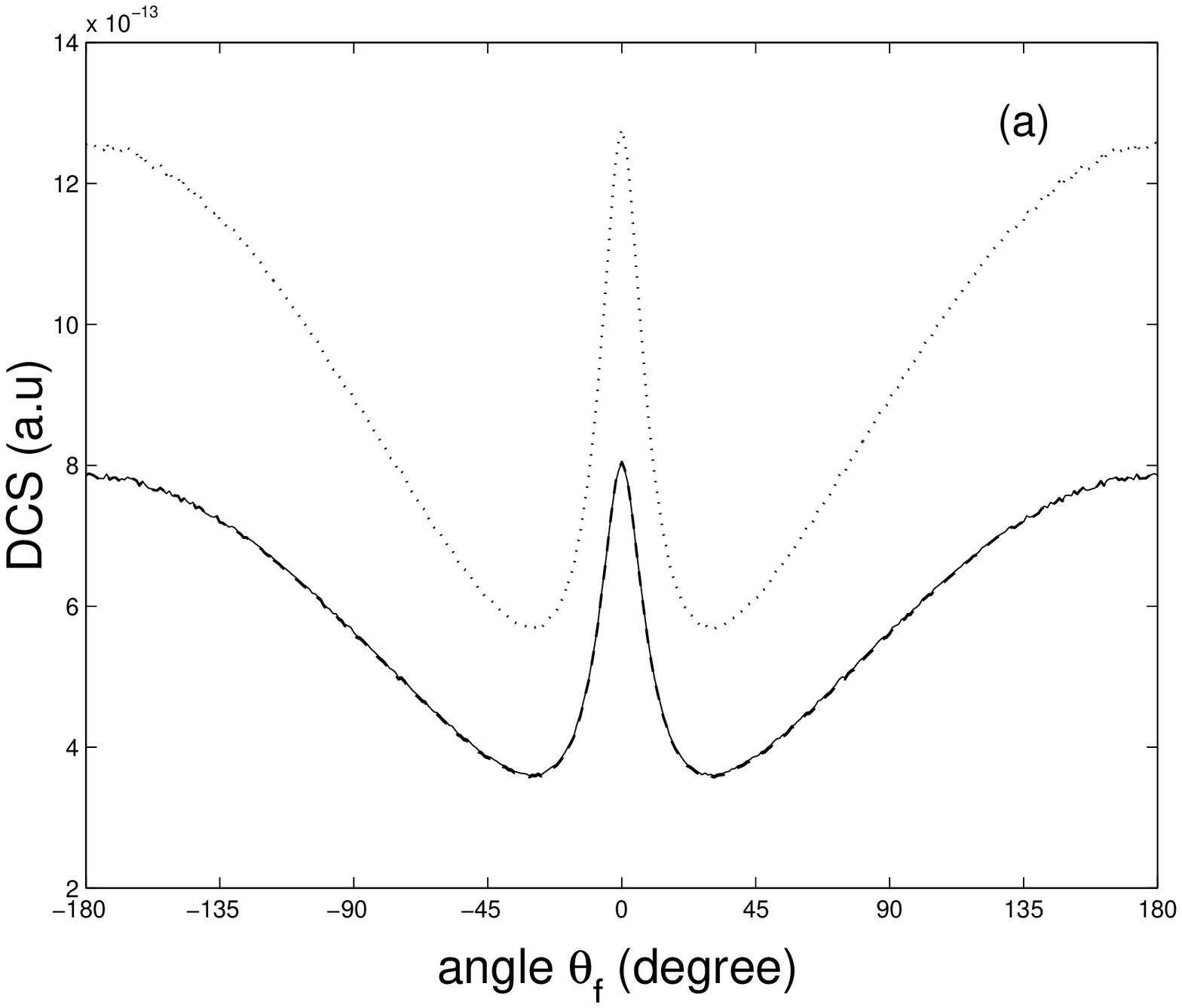}
\includegraphics[angle=0,width=3 in,height=2.17 in]{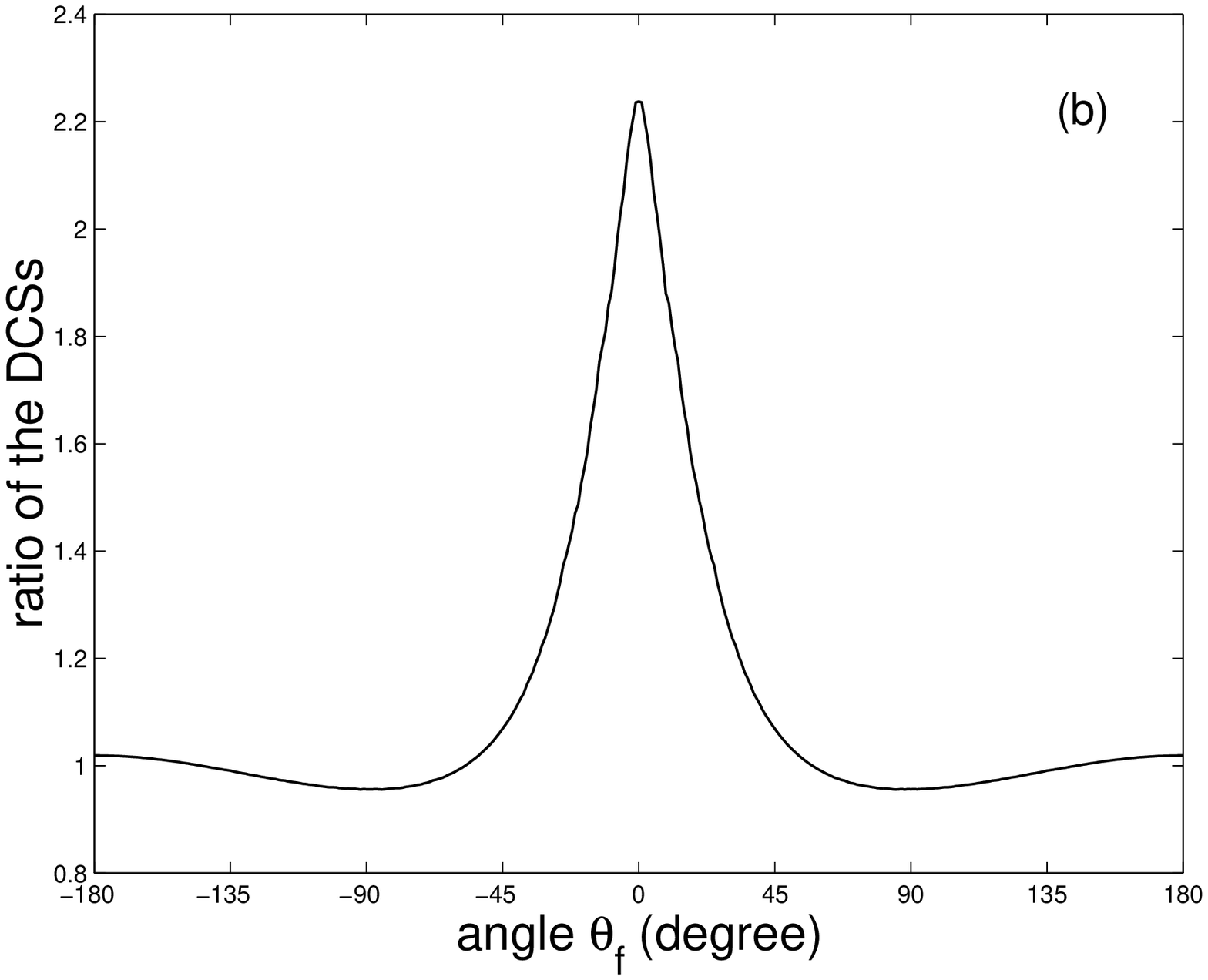}
\caption{(a) Summed differential cross sections $d\sigma/d\Omega$
of $\pm 100$ peaks around the elastic one as a function of the
angle $\theta_f$ for a relativistic parameter $\gamma=2.00$ and
an electric field strength $\varepsilon=1.00\,a.u$. The solid
line denotes the result for the Dirac-Volkov electrons, the long
dashed one sketches the values for $(d\sigma/d\Omega)^{[1]}_{DV}$
and the short dashed is the result for spinless particles. (b):
Ratio $R$ of the two Dirac-Volkov DCSs for a relativistic
parameter $\gamma=2.00\,a.u$ and an electric field strength
$\varepsilon=5.89\,a.u$}
\end{figure}
\subsection{Relativistic and high electric field strength regime}
To study this regime, we use the same parameters as in \cite{1},
that is a relativistic parameter $\gamma =5$ or an incoming
electron kinetic energy $T_i=4c^2\,a.u=2.045\,Mev$. The electric
field strength is $\varepsilon=5.89 \,a.u$. In
table(\ref{table4}), some cosines of the angles
$cos(\widehat{\mathbf{q_{i}, \mathbf{q_f}}})$ and
$cos(\widehat{\mathbf{p_{i}, \mathbf{p_f}}})$ are given. In this
regime, the dressing of the electron angular coordinates is very
important.
\begin{center}
\begin{tabular}{|c|c|}
\hline\hline
$\cos(\widehat{\mathbf{p}_i,\mathbf{p}_f})$&$\cos(\widehat{\mathbf{q}_i,\mathbf{q}_f})$\\
\hline\hline
0.853553&0.866661\\
\hline\hline
0.850868&0.861677\\
\hline\hline
0.847923&0.856549\\
\hline\hline
0.844719&0.851270\\
\hline\hline
 0.841259&0.845833\\
 \hline\hline
\end{tabular}
\begin{table}[h]
\caption{\label{table4}}
\end{table}
\end{center}
\indent In the upper part (a) of Figure 6, we show the various
DCSs. \begin{figure}[ht] \centering
\includegraphics[angle=0,width=3 in,height=2.3 in]{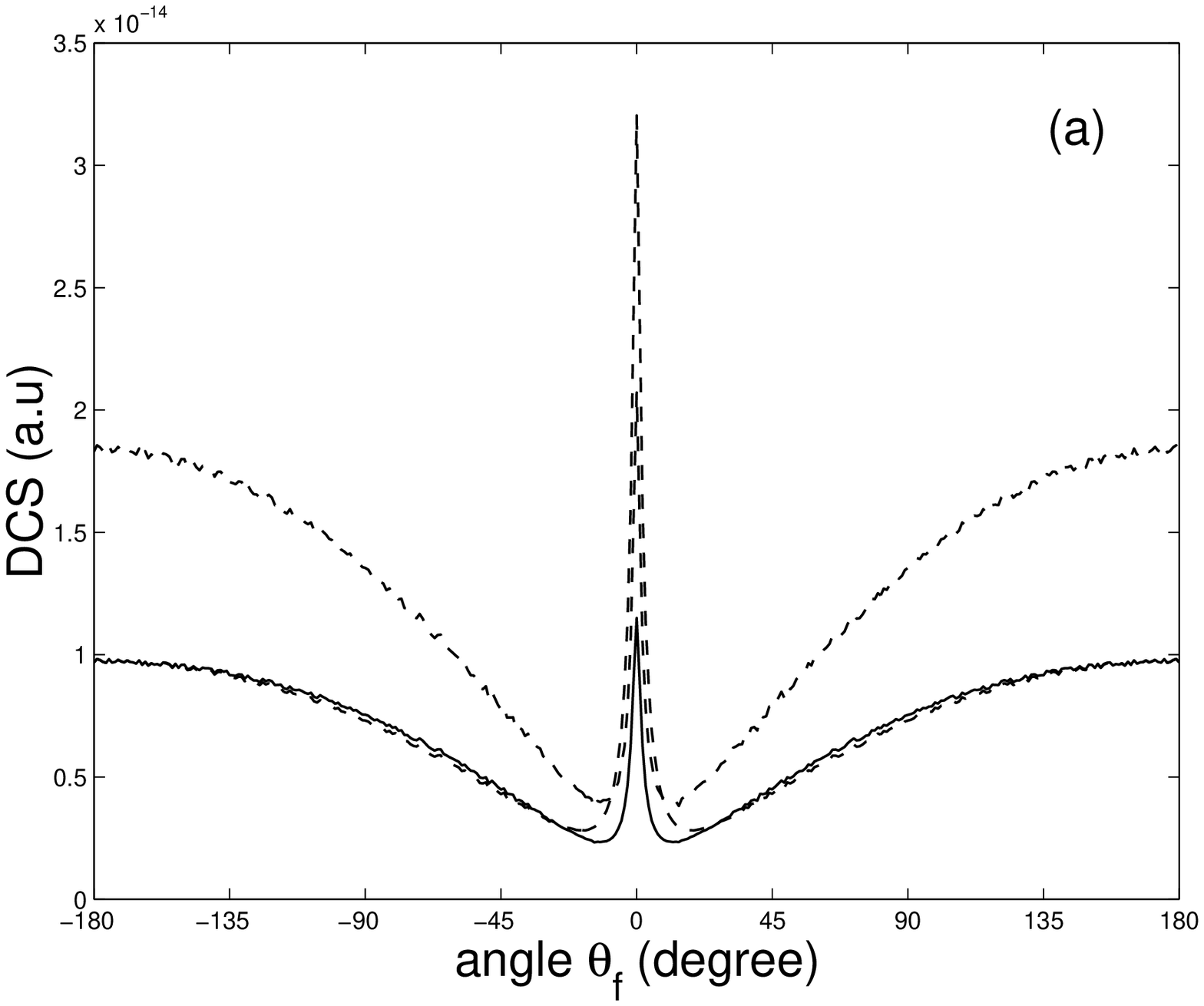}
\includegraphics[angle=0,width=3 in,height=2.25 in]{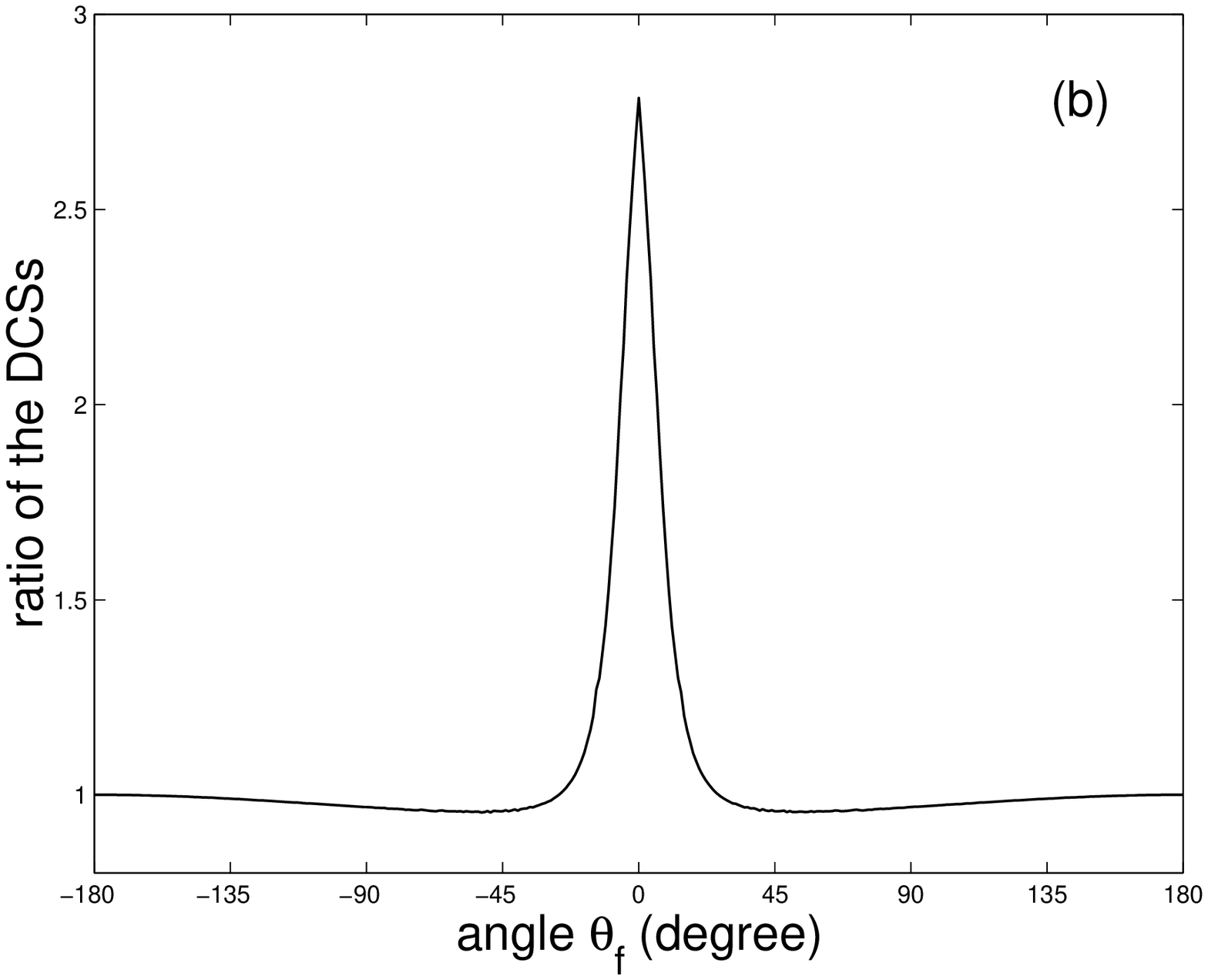}
\caption{(a): Summed differential cross section $d\sigma/d\Omega$
of $\pm 100$ peaks around the elastic one as a function of the
angle $\theta_{f}$ for a relativistic parameter $\gamma=5.00\,a.u$
and an elastic field strength $\mathbf{\varepsilon}=5.89\, a.u$.
Th solid line denotes the result for Dirac-Volkov electrons, the
long dashed one sketches the values for
${(d\sigma/d\Omega)_{DV}^{\cite{1}}}$ and the short dashed is the
result for spinless particles. (b): Ratio $R$ of the two
Dirac-Volkov DCSs for the same values of the relevant parameters,
$\gamma=5.00\,a.u$ and $\mathbf{\varepsilon}=5.89\,a.u$}.
\end{figure}
For angles $  \theta_{f}=\pm 180^{\circ}$, the agreement
between our results and the results of \cite{1} is good but
deteriorates for small values of $\theta_{f}$. For
$\theta_{f}=0^{\circ}$, the result of our work gives a value (
scaled in $ 10^{-14}\, a.u $ ) $(\frac
{d\sigma}{d\Omega_{f}})_{DV}=1.15 $ while de corresponding result
found using Eq.(26) of \cite{1} is $(\frac
{d\sigma}{d\Omega_{f}})_{DV}^{\cite {1}}=3.204 $ . Our results
(solid line) are always smaller than the results for spinless
particles while those obtained using Eq.(26) of \cite{1} are
greater than $(\frac{d\sigma}{d\Omega_{f}})_{KG}$ for small
angles around the direction of the laser propagation. In the lower
part (b) of Figure 6, we show the ratio R defined by
Eq.(\ref{62}). For $\theta_{f}=0^{\circ}$, this ratio is
$R=2.787$.

\section{Conclusion}
In this work, we derived the correct expression of the first Born
differential cross section for the scattering of the Dirac-Volkov
electron by a Coulomb potential of a nucleus in the presence of a
strong laser field. We have given the correct relativistic
generalization of the Bunkin and Fedorov treatment \cite{7} that
is valid for an arbitrary geometry. We are adamant that the core
of the whole controversy stems from the fact that in \cite{1},
the vector $\eta^{\mu}$ introduced in Eq.(\ref{43}) of our work
has not been properly dealt with while it is the common method to
use when a trace contains a $\gamma^0$ matrix. Any standard QED
textbook introduces this very elementary method. Comparison of
our numerical calculations \cite{9} with those of Szymanowski et
al. \cite{1} shows qualitative and quantitative differences when
the incoming total electron energy and the electric field
strength are increased particularly in the direction of the laser
propagation. The difference between our results and those of
\cite{1} can only be traced back to the mistakes and the omitted
term in Eq.(26) of \cite{1}. The corrections that we made allowed
us to study other processes that were published in Physical
Review A., namely an first article concerning the relativistic
electronic dressing in laser assisted electron-hydrogen elastic
collisions \cite{10}, another concerning the process of Mott
scattering in an elliptically polarized laser field \cite{11} as
well as a third work dealing with the process of Mott scattering
of polarized electrons in a strong laser field \cite{12}. For the
difficult process of ionization of atomic hydrogen by electron
impact, we published an article concerning the importance of the
relativistic electronic dressing in laser-assisted ionization of
atomic hydrogen by electron impact \cite{13}. All these works
relied heavily on the corrections that we made in this work.

\end{document}